\documentclass[11pt]{article}
\usepackage[left=1in,top=1in,right=1in,bottom=1in,head=0in,nofoot]{geometry}

\usepackage{setspace,url,bm,amsmath} 
\usepackage{scalerel}

\usepackage{xcolor}
 \usepackage{multirow}
 \usepackage{arydshln}
 \usepackage{mathtools}
\usepackage{titlesec} 
\titlelabel{\thetitle.\quad} 
\titleformat*{\section}{\bf\Large\center}

\usepackage{authblk}
\usepackage{float}
\usepackage{graphicx} 
\usepackage{bbm}
\usepackage{latexsym}
\usepackage{caption}
\usepackage[margin=20pt]{subcaption}
\usepackage{enumerate}
\graphicspath{ {./Graphs/} }

\usepackage{tcolorbox}

\usepackage{amsthm}
\usepackage{amssymb}
\usepackage{amsmath}
\usepackage{color}

\usepackage{comment}
\theoremstyle{definition}
\newtheorem{assumption}{Assumption}
\newtheorem*{theorem*}{Theorem}

\newtheorem*{rmk*}{remark}

\newtheorem{example}{Example}

\newtheorem{remark}{Remark}

\newtheorem*{corollary*}{Corollary}

\usepackage{natbib} 
\bibpunct{(}{)}{;}{a}{}{,} 

\usepackage{etoolbox} 
\apptocmd{\sloppy}{\hbadness 10000\relax}{}{} 

\usepackage{multibib}
\newcites{sec}{References}

\usepackage{color}
\usepackage{listings}
\usepackage{hyperref}
\usepackage{booktabs}

\usepackage{lscape}

\RequirePackage[normalem]{ulem}

\def \P{\mathbb{P}}

\def \E {\mathbb{E}}

\newcommand{\indep}{\perp \!\!\! \perp}

\usepackage{wasysym}

\allowdisplaybreaks

\begin{document}

\singlespacing

\title{\bf A Distributional Perspective on Pearl's Causal Hierarchy: From Marginal to Joint and Individualized Potential Outcomes}

\author[1]{Peng Wu}
\author[2]{Linbo Wang\thanks{Corresponding author: linbo.wang@utoronto.ca}}
\affil[1]{\small School of Mathematics and Statistics, Beijing Technology and Business University,  China} 
\affil[2]{\small Department of Statistical Sciences, University of Toronto, Toronto, ON, Canada}


\date{}

\maketitle

\begin{abstract}
Pearl's causal hierarchy is a foundational lens for formulating causal questions and is most often discussed within the framework of structural causal models. We recast the hierarchy in potential outcomes language and make its information ordering operational at the level of causal estimands. Specifically, we classify an estimand according to whether it requires marginal potential outcome distributions, their joint distribution or nested cross-world quantities, or individual-level counterfactual outcomes. We apply this criterion systematically to a broad range of estimands, including cases whose classification depends on the formulation of the scientific question. We then clarify what additional assumptions are needed when an estimand depends not only on the marginal distributions of potential outcomes, but also on their unobserved joint distribution. In particular, randomization identifies the marginals, whereas monotonicity, copula restrictions, rank preservation, and partial identification restrict or characterize the remaining uncertainty about the joint distribution. The resulting framework provides a practical map from a scientific question to an estimand, the probabilistic object it requires, and the assumptions needed for identification.\end{abstract}




\onehalfspacing


\section{Introduction}
\label{intro} 
Understanding causal relationships is a fundamental goal across a wide range of domains and has gained increasing attention in both academic and industry communities in recent years~\citep{Imbens-Rubin2015, Pearl2019, Hernan-Robins2020}.  
Pearl articulates a three-layer causal hierarchy~\citep{Shpitser-Pearl2008, pearl2009causality, Pearl-Mackenzie2018}--commonly referred to as the ``Ladder of Causation''--that organizes causal reasoning into association, intervention, and counterfactuals (See Table \ref{tab-a1} for details). This hierarchy distinguishes classes of causal queries with progressively greater conceptual and inferential demands, and has become a foundational lens for understanding the scope of causal analysis~\citep{Bareinboim-etal2022}.
In addition, several studies have examined this hierarchy from complementary perspectives grounded in the structural causal model (SCM) framework, including logical-probabilistic, inferential-graphical, and computational complexity viewpoints~\citep{Bareinboim-etal2022, Julian-etal2025}. 

\medskip 
{\bf Motivation.} 
Existing discussions are primarily formulated in structural causal model language and are not typically organized around an explicit estimand-level classification rule stated in potential outcomes terms. We develop and systematically apply such a rule to address the question below. 

\begin{tcolorbox}
Q1: How can we determine which layer of causation a given causal estimand belongs to, and how can diverse causal estimands be systematically classified? 
\end{tcolorbox}


\begin{table}[t]
\centering
\caption{Pearl's causal hierarchy~\citep{Pearl2019}.
Questions at layer $l$ ($l = 1, 2, 3$) can be answered only if information from layer $l$ or higher is available. Here, $A$ denotes the treatment (intervention), $X$ the covariates, and $Y$ the outcome.}
\resizebox{1\columnwidth}{!}{
\begin{tabular}{lcc} 
\toprule 
Layer & Typical Activity & Typical Questions  \\ \midrule
 1. Association $\P(Y \mid X)$ & Seeing & What is? How would seeing $A$ change my belief in $Y$? \\  
2. Intervention $\P(Y\mid do(A=a), X)$  & Doing. Intervening   & What if? What if I do $A$.  \\
  3. Counterfactuals $\P(Y(a) \mid A=a', Y(a'))$  & Imagining. Retrospection   & Why? Was it $A$ that caused $Y$? what if I had acted differently? \\ 
\bottomrule 
\end{tabular}}
\label{tab-a1}
\end{table}

\begin{figure*}[t]
    \centering
    \includegraphics[width=0.9\textwidth]{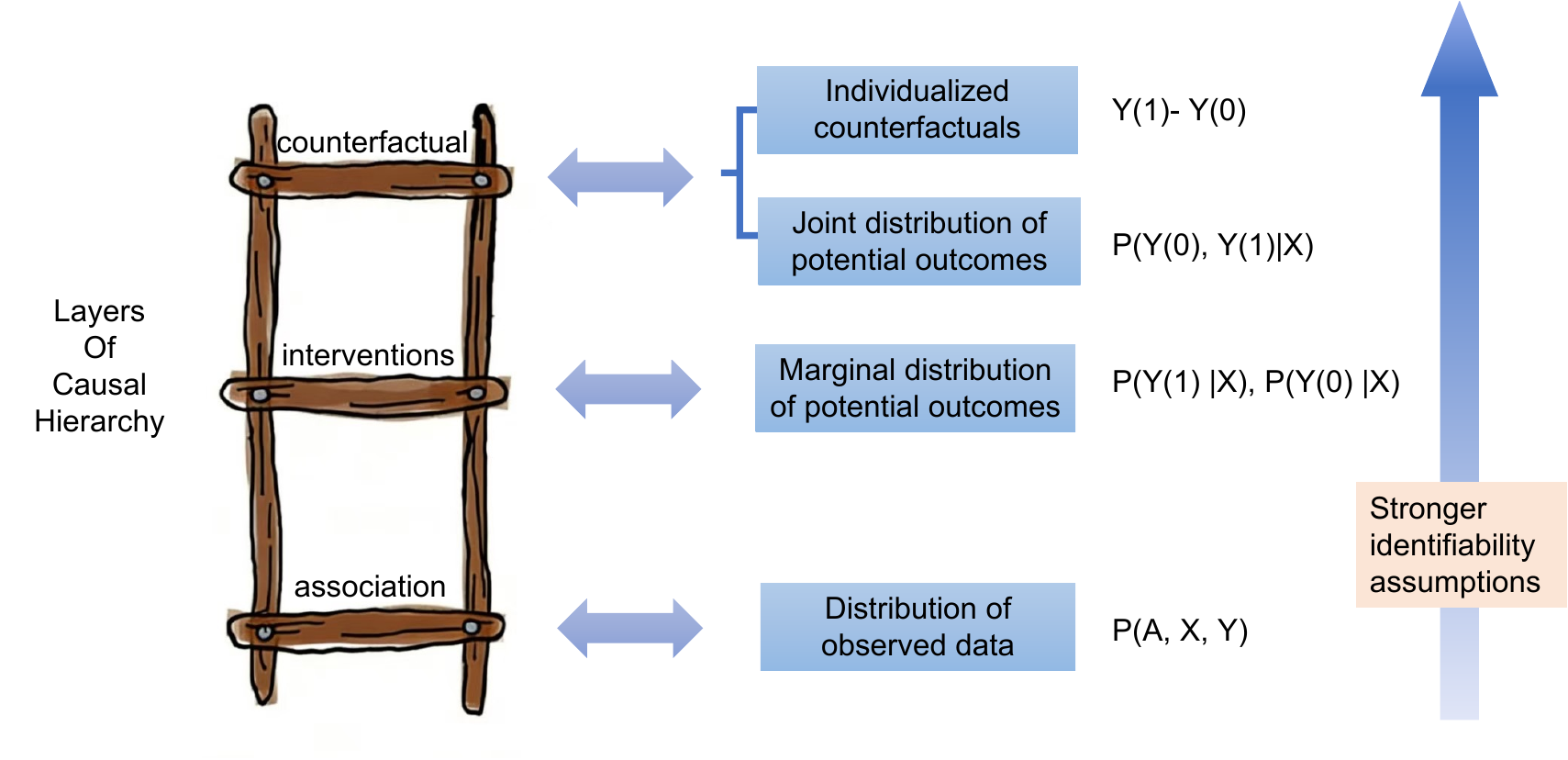}
    \caption{
    The proposed potential outcomes perspective {on Pearl's causal hierarchy}.}
    \label{fig1}
    \vskip -6pt
\end{figure*}

{\bf Our Perspective.} 
We address Q1 using an operational criterion based on the probabilistic object required by an estimand. Although the underlying ordering of marginal, joint, and individual-level information is familiar, stating the criterion at the estimand level makes it possible to classify causal targets directly in potential outcomes language. Our answer to Q1 is as follows:

\begin{tcolorbox}
\centerline{\bf Potential Outcomes Perspective} 

{\bf Second layer: Intervention}.
A causal estimand that depends only on the {\bf marginal} distributions of potential outcomes (conditional on observed treatment and covariates) belongs to the second layer of causation.

\textbf{Third layer: Counterfactuals}. A causal estimand that further depends on the {\bf joint} distribution of potential outcomes
 (conditional on observed treatment and covariates), involves nested potential outcomes under different, mutually exclusive interventions (a.k.a. cross-world relationships), or targets potential outcomes at the individual level belongs to the third layer of causation. 
\end{tcolorbox}

We omit discussion of the first layer of causation, as causal inference primarily focuses on the second and third layers~\citep{Pearl2019}.  Figure~\ref{fig1} illustrates the connection between the causal hierarchy and potential outcomes; see Section~\ref{Sec3-1} for details. The resulting classification is consistent with the familiar information ordering across the hierarchy: moving from marginal distributions to joint distributions and ultimately to individualized potential outcomes generally requires progressively stronger identifiability assumptions.

\medskip 
\textbf{Contributions.} 
Rather than proposing a new causal hierarchy, this paper contributes an operational taxonomy and synthesis within the potential outcomes framework. First, we provide an estimand-level diagnostic: a target is classified by whether it requires marginal potential outcome distributions, a joint distribution or nested cross-world quantity, or individual-level counterfactual outcomes. This turns the familiar information ordering into a rule that can be applied directly to causal estimands.

Second, we apply the diagnostic systematically to a broad range of targets, including ambiguous cases such as the average treatment effect on the treated and scientifically distinct formulations of mediation effects. We also distinguish cross-world targets from individual-level counterfactual targets within the third layer; see Table~\ref{tab1} and Examples~\ref{example1}--\ref{example10}. This classification clarifies what a scientific question asks for and prevents a population-average estimand from being used to support an individual-level claim~\citep{Lei-Candes2021, 2022nathan}.

Third, we clarify what additional assumptions are needed when an estimand depends not only on the marginal distributions of potential outcomes, but also on their unobserved joint distribution. Randomization identifies the marginals, while monotonicity, copula restrictions, rank preservation, and partial identification restrict or characterize the remaining uncertainty about the joint distribution. This perspective links each scientific question to its estimand, required probabilistic object, and identifying assumptions, and thereby clarifies when an assumption set is insufficient or unnecessarily restrictive.
\section{Preliminaries}
\label{setup}

\subsection{Notation}
Let $A\in \mathcal{A}$ denote the treatment (or intervention), $X\in \mathcal{X}$ a vector of pre-treatment covariates, and $Y\in \mathcal{Y}$ an outcome variable. 
Under the potential outcomes framework, let $Y(a)$ denote the potential outcome that would be observed if the treatment were set to $A=a$.
Under the stable unit treatment value assumption, i.e., there are no multiple versions of treatment and no interference among individuals, the observed outcome is  $Y=Y(A)$. 
Let $\mathbb{P}$ denote the target population of interest, with $\mathbb{E}$ the corresponding expectation operator. 

\begin{figure}[ht]
    \centering
    \includegraphics[width=0.5\textwidth]{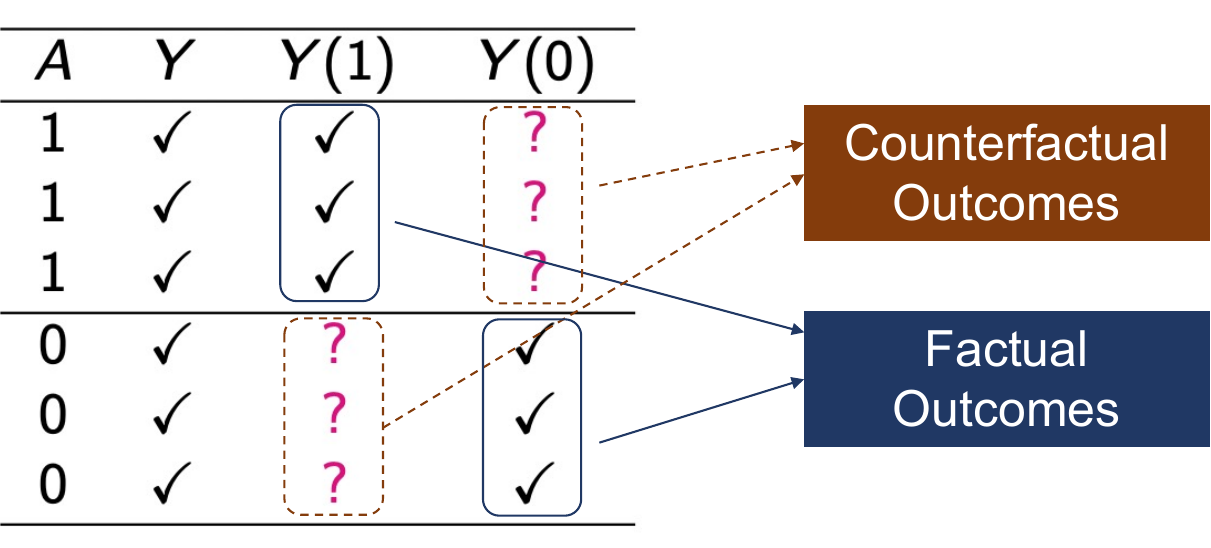}
        \caption{
        Illustration of factual and counterfactual outcomes under binary treatment, where $\checkmark$ and $?$ mean observed and unobserved, respectively. We omit $X$ for simplicity.}
    \label{fig1-appendix}
\end{figure}

With this notation in place, we distinguish between factual and counterfactual outcomes. Potential outcomes are defined prior to measurement; after measurement, one potential outcome becomes factual while the others remain counterfactual~\citep{wang2025causalinferencetaleframeworks}. Specifically, for an individual with $A=a$, $Y(a)$ is the factual (observed) outcome, whereas $Y(a')$ for $a' \neq a$ is the counterfactual (unobserved) outcome; see Figure~\ref{fig1-appendix}  for illustration. 


In Section~\ref{Sec3-5}, we introduce a post-treatment variable observed after treatment assignment and before the outcome. This facilitates discussion of long-term causal effects, mediation analysis, counterfactual fairness metrics, and principal causal effects, demonstrating the applicability of our proposed perspective and classification.

\subsection{Pearl's Causal Hierarchy and Structural Causal Model}

We provide a brief introduction to Pearl’s causal hierarchy and the definition of the structural causal model.

\medskip 
{\bf Pearl's Causal Hierarchy}.  We present Pearl's causal hierarchy in Table~\ref{tab-a1}  and highlight two key points. First, interventional queries (the second layer) are typically prospective, as they reason about the effects of hypothetical interventions on future outcomes based on current observations, whereas counterfactual queries (the third layer) are  retrospective, as they consider alternative past outcomes for events that have already occurred~\citep{Pearl2019}.
Second, questions at layer $l$ ($l = 1, 2, 3$) can be answered only if information from layer $l'$ with $l' \geq l$ is available~\cite{Ibeling-2020-AAAI, Bareinboim-etal2022}. 
See \citet{Pearl2019} for more details. 



\medskip 
{\bf Structural Causal Model} (SCM,  \citeauthor{pearl2009causality}, \citeyear{pearl2009causality}). An SCM $\mathcal{M}$ consists of a graph $\mathcal{G}$ and a corresponding set of structural equation models $\mathcal{F} = \{f_1, ..., f_p \}$.  
The nodes in $\mathcal{G}$ are divided into two categories: (a)  exogenous variables ${\bf U} = (U_1, ..., U_p)$,  which represent the environment during data generation, assumed to be mutually independent;    
(b) endogenous variables $\textbf{V} = \{V_1, ...,  V_p\}$, which denote the relevant features that we need to model in a question of interest. For a variable $V_j$, its value is determined by a structural equation $V_j = f_j(PA_j, U_j), ~j=1, ..., p$, where $PA_j$ stands for the set of parents of $V_j$. 
The SCM provides a formal language for describing how the variables interact and how the resulting distribution would change in response to certain interventions. 

\medskip 
The SCM will be used to characterize counterfactual outcomes in Section~\ref{Sec3-4}. For a detailed discussion of how SCMs can be formulated in terms of potential outcomes and how the two frameworks can be translated into one another, we refer the reader to \citet{RichardsonRobins2013} and \cite{wang2025causalinferencetaleframeworks}. 



\section{A Potential Outcomes Perspective}  \label{Sec3}
In this section, we provide a detailed exposition of the potential outcomes perspective on Pearl's causal hierarchy. Section~\ref{Sec3-1} presents an overview, while Sections~\ref{Sec3-2}--\ref{Sec3-4} focus on the second and third layers of causation, offering a classification of various causal estimands and highlighting key identifiability challenges and associated identifiability strategies across layers. Section~\ref{Sec3-5} extends these discussions to settings involving post-treatment variables. Section~\ref{Sec3-6} discusses the challenges and limitations of third-layer estimands.

\subsection{Overview}  \label{Sec3-1}

Figure~\ref{fig1} illustrates the proposed potential outcomes perspective and its correspondence with Pearl's causal hierarchy. We briefly describe this perspective below.
     
\medskip 
\textbf{Second layer: Intervention.} 
Causal estimands at this layer are functionals of the marginal distributions of potential outcomes. In other words, each potential outcome corresponds to the outcome in an intervention world in which all individuals receive the same treatment, and second-layer estimands depend only on one or more marginal potential outcome distributions under a single intervention. A causal query (or estimand) that can be decomposed into one or more sub-queries, each involving a single intervention, belongs to the second layer. Treatment randomization identifies the marginal distributions of potential outcomes and is therefore sufficient to identify estimands at this layer. 
Typical causal estimands at this level include the average treatment effect, the quantile treatment effect, and the dose-response function. See Section~\ref{Sec3-2} and Table~\ref{tab1} for details.  

\medskip 
\textbf{Third layer: Counterfactuals.}  
We divide this layer into two sublayers by inferential complexity. 

 
\emph{Sublayer 1: Cross-World Causal Queries.} 
Causal estimands at this layer depend on features of the joint distribution of potential outcomes, or equivalently on nested potential outcomes defined under multiple, mutually exclusive interventions. Such estimands cannot be expressed as functionals of marginal potential outcome distributions alone. 
Because they simultaneously reference outcomes from different interventional worlds, we refer to these estimands as cross-world causal queries. 
Treatment randomization alone identifies only marginal potential outcome distributions and is therefore insufficient for identifying estimands at this layer. 
Typical estimands at this layer include attribution metrics, benefit and harm rates, persuasion rate, and natural direct and indirect effects. These estimands are useful when the scientific question intrinsically concerns attribution, the joint distribution of benefits and harms, or another cross-world comparison. They can complement second-layer averages by describing subgroup-level response distributions, but they need not improve a decision when the scientific question can be formulated using marginal potential outcome distributions alone. See Section~\ref{Sec3-3} and Table~\ref{tab1} for more details.

\medskip 
\emph{Sublayer 2: Individual-Level Counterfactual Queries.}
This layer focuses on individualized treatment effects (ITEs). Learning ITEs requires much stronger identifiability assumptions than those needed for estimands in other layers.

\begin{remark}
The classification is consistent with a familiar monotonic relationship: knowing individual-level counterfactual outcomes implies knowledge of the joint distribution of potential outcomes, and knowing the joint distribution implies knowledge of the marginal distributions--but not vice versa. We use this relationship as background for the estimand-level diagnostic, rather than as a new property of Pearl's hierarchy.
\end{remark}
%


In this paper, we primarily focus on the third layer consisting of two sublayers, and only briefly discuss the second layer, as it has been extensively studied in the existing literature and textbooks~\citep{Imbens-Rubin2015, Rosenbaum2020, Hernan-Robins2020, Ding2023}. 
In Sections~\ref{Sec3-2}--\ref{Sec3-5}, we omit the covariates $X$ for notational simplicity unless explicitly stated. 
The definitions below may be read conditionally on $X=x$. The corresponding identification results require the conditional versions of the stated assumptions, together with appropriate positivity and regularity conditions.

\begin{table}[t]
\centering
\caption{Classification of various estimands and their associated identifiability strategies. Note: this table lists only a subset of typical causal estimands and is not intended to be exhaustive.} 
\vspace{2pt}
\renewcommand{\arraystretch}{1.1}  
\resizebox{1 \linewidth}{!}{
\begin{tabular}{clcc} 
\toprule 
Layer & Typical Estimands (or Classes Thereof) & Main Challenges & (Partial) Identifiability Strategies   \\ \midrule
  & Average Treatment Effect, Causal Risk and Odd Ratios (Example \ref{example1}) &        & Ignorability (or Randomization)   \\ 
 Second Layer  & Quantile and Distributional Treatment Effects (Example \ref{example1}) & (Unmeasured) Confounding   & Auxiliary Variable (e.g., IV) \\ 
(Marginal Distribution)  & Dose-Response Function, Average Derivative Effect (Example \ref{example2}) & Between Treatment and Outcome &  Data Fusion \\ 
     &  Short-Term and Long-Term Treatment Effects (Example \ref{example8}) &  &  Sensitivity Analysis \\
    &      Counterfactual Parity and Total Effect (Examples \ref{example9} and \ref{example10}) \\  
     \midrule
    &  Probability of Causation (Example \ref{example3})   &     & Ignorability  + Independence \\
   & Treatment Benefit and Harm Rates (Example \ref{example4})   & (Unmeasured) Confounding  & Ignorability  + Monotonicity   \\
Third Layer    & Effect of Persuasion (Example \ref{example5})   & Between Treatment and Outcome &   Ignorability  + Association Parameter  \\
  (Joint Distribution or   & Distribution of the ITE (Example \ref{example6})   & $+$  & Ignorability  + Copula Model  \\ 
    Nested Potential Outcomes in  & Principal Causal Effects (Example \ref{example7})  & Dependence Between   &   Data Fusion \\
     Cross Worlds)  & Principal Fairness (Example \ref{example9})  & Cross-World Potential Outcomes &  Sensitivity Analysis Under Ignorability  \\ 
        & Natural Direct and Indirect Effects (Example \ref{example10}) &   &  Partial Identification Under Ignorability  \\ \midrule
  Third layer     &   Individualized Treatment Effect   &  Connection Between Factual  and   &   Abduction-Action-Prediction   \\ 
    (Individual Level)     &  Counterfactual Fairness (Example \ref{example9})  &
     Counterfactual Outcomes at the  & Rank Preservation    \\
        &     &
    Individual Level &   Conformal Inference    \\
\bottomrule 
\end{tabular}}
\label{tab1}
\par\vspace{2pt}
{\footnotesize 
Note: The notation ``$A + B$'' denotes $A$ and $B$; that is, both $A$ and $B$ are required or exist.}
\end{table}

\subsection{Second Layer: Intervention}  \label{Sec3-2}


 We introduce typical causal estimands at the second layer of causation.

\begin{example}[Binary Treatment, Second Layer] \label{example1} \text{   }

 $\bullet$ Average treatment effect (ATE):  $\textup{ATE}  =  \E[ Y(1) - Y(0)  ].$

 $\bullet$ ATE on treated (ATT): $\textup{ATT} = \E[ Y(1) - Y(0)   \mid A=1 ].$ 

The classification of the ATT depends on the convention used for the hierarchy. Under Pearl's syntactic convention, the counterfactual component $\E\{Y(0)\mid A=1\}$ combines a potential outcome under an intervention with conditioning on the observed treatment and can therefore be represented as a third-layer query. This syntactic classification is distinct from identification: under randomization the ATT coincides with the ATE in the target population, and under conditional exchangeability and positivity it is identifiable by standard adjustment. Under the distributional convention adopted here, we classify the ATT as a second-layer estimand because it depends on the two marginal distributions of $Y(1)$ and $Y(0)$ within the treated subpopulation, rather than on their joint distribution or cross-world dependence. Other conventions can therefore assign the ATT to a different layer without implying a disagreement about its empirical identification.

 $\bullet$ Quantile treatment effect (QTE, \citeauthor{Firpo2007}, \citeyear{Firpo2007}): 
	$\textup{QTE}_{\tau}  = q_{1,\tau}  - q_{0, \tau},$
where $q_{a, \tau} = \inf\{q:  \P( Y(a)\leq q ) \geq \tau \}$ is the $\tau$-quantile of the distribution of $Y(a)$ for $a = 0, 1$.

 $\bullet$ Distributional treatment effect (DTE, \citeauthor{Oka-etal2025}, \citeyear{Oka-etal2025}): 
 $\textup{DTE}(y) =  F_{Y(1)}(y) -  F_{Y(0)}(y),$  
where $F_{Y(a)}(\cdot)$ is the cumulative distribution of $Y(a)$ for $a = 0, 1$, and $y \in \mathcal{Y}$.

 $\bullet$ In addition, for binary outcomes, we usually define {causal} risk ratio (CRR) and {causal} odd ratio (COR): 
    \[   \textup{CRR} = \frac{\P(Y(1) = 1)}{ \P(Y(0) = 1) },  \quad  \textup{COR} = \frac{\P(Y(1) = 1)/\P(Y(1) = 0)}{ \P(Y(0) = 1)/\P(Y(0) = 0)}.  \]
\end{example}


\begin{example}[Continuous Treatment, Second Layer] \label{example2}
\text{   }

 $\bullet$ Effect curve or dose–response function (DRF)~\citep{Galvao02102015, Kennedy-etal2017}: $\textup{DRF}(a) = \E[Y(a)].$ 

 $\bullet$ Average derivative effect (ADE,  
 \citeauthor{newey1993efficiency}, \citeyear{newey1993efficiency}; \citeauthor{dong2025marginalcausaleffectestimation}, \citeyear{dong2025marginalcausaleffectestimation}): $$\textup{ADE}(a) =\frac{\partial \E[Y(a)] }{ \partial a}.$$ 
\end{example}



For identifying estimands at the second layer, the primary challenge arises from confounding between treatment and outcomes. The most commonly used identifiability conditions are ignorability ($A \indep Y(a)$), also referred to as no unmeasured confounding~\citep{rosenbaum1983central}, and the overlap condition ($0 < \Pr(A=1) < 1$), which is satisfied under a well-designed randomized treatment assignment.
When randomization is infeasible and unmeasured confounding is a concern, researchers may pursue alternative strategies. These include leveraging auxiliary variables, such as instrumental variables \citep{Imbens2004,wang2018bounded} and negative controls \citep{Lipsitch-etal2010, Hu-etal2023-NegativeControl}; adopting data-fusion approaches that combine multiple complementary data sources \citep{Colnet-etal2024, Wu-etal-2025-Compare}; and exploiting structural restrictions in multi-dimensional treatments and/or outcomes \citep{zhou2024promises,tang2026synthetic}. In addition, sensitivity analysis provides a fundamental tool for assessing the robustness of causal conclusions to violations of ignorability \citep{Kallus-Zhou2018, Rosenbaum2020, Ding-etal2022, Ding2023}.

\subsection{Third Layer: Cross-World Causal Queries}  \label{Sec3-3}

In this subsection, we first present various causal estimands involving the joint distribution of potential outcomes, and then discuss the key identifiability challenges and review the associated identification strategies.

\subsubsection{Causal Estimands}
 We present representative third-layer estimands involving joint distributions in the following examples, each illustrating an interesting application scenario.

\begin{example}[Probability of Causation] \label{example3}
Causal inference concerns not only the effects of causes but also the causes of observed effects~\citep{Dawid2022}, the latter often referred to as attribution analysis~\citep{pearl2009causality, Pearl-etal2016-primer}. For binary treatment and outcomes, two standard attribution measures are the probability of necessary causation (PN) and the probability of sufficient causation (PS), defined as follows \citep{pearl1999, Tian-Wu2025}: 
\begin{align*}  
\textup{PN}(A \Rightarrow Y)  ={}& \P(  Y(0) = 0 \mid A=1, Y=1 ),\\
\textup{PS}(A \Rightarrow Y)  ={}& \P( Y(1)= 1 \mid A = 0, Y = 0 ). 
\end{align*}
Observe that we can rewrite PN and PS as
$
\textup{PN}(A \Rightarrow Y)
    = \P\!\left(Y(0)=0 \mid A=1, Y(1)=1\right)
$
and
$
\textup{PS}(A \Rightarrow Y)
    = \P\!\left(Y(1)=1 \mid A=0, Y(0)=0\right).
$
These expressions make it explicit that PN and PS depend on the joint distribution of the potential outcomes $(Y(0),Y(1))$.

The PN provides a probabilistic formalization of the but-for principle in legal reasoning, while PS captures a complementary notion of causal sufficiency~\citep{Pearl-Mackenzie2018}.
In recent years, several studies have extended PN and PS to non-binary treatments, multiple binary treatments, and non-binary outcomes; see \citet{lu2023evaluating, zhao2023conditional, li2024probabilities, li2024retrospective, Zhang-etal2025, Luo-etal-2025} for more details. 
\end{example}

\begin{example}[Treatment Benefit and Harm Rates] \label{example4} 
For a binary treatment, assuming larger outcomes are preferable, the treatment harm rate (THR) and treatment benefit rate (TBR)~\citep{Gadbury2004, Zhang-etal2013, Huang-etal2012, Mueller-Pearl2023, MuellerPearl2023-CausalInference, shen2013treatment, Yin-etal2018, 2022nathan, li2023trustworthy, Wu-etal-2024-Harm} are defined as:  
  \begin{align*}
      \textup{THR} ={}& \P(Y(1) - Y(0) < 0), \\
       \textup{TBR} ={}& \P(Y(1) - Y(0) > 0). 
  \end{align*} 
For discrete outcomes, the treatment no-effect rate (TNR) completes this decomposition:
\[
\textup{TNR}=\P\{Y(1)-Y(0)=0\}=1-\textup{THR}-\textup{TBR}.
\]
Although the TNR is not the primary focus in applications centered on benefit and harm, it records the proportion whose outcome would be unchanged by treatment.
These quantities measure the proportions of individuals who experience better or worse outcomes under treatment compared to control, respectively. In addition, one may define the treatment harm quantity (THQ) as
$$\textup{THQ} = \E[ (Y(0) - Y(1)) \mathbb{I}(Y(1)-Y(0)<0)],$$
which quantifies the magnitude of treatment-induced harm. The conditional average treatment effect (CATE),
$\tau(x)=\E\{Y(1)\mid X=x\}-\E\{Y(0)\mid X=x\}$, depends only on the marginal distributions of $Y(1)$ and $Y(0)$ given $X=x$ and is therefore a second-layer estimand. For binary outcomes, define the conditional treatment benefit, harm, and no-effect rates as
\[
\begin{split}
\textup{TBR}(x)&=\P\{Y(0)=0,Y(1)=1\mid X=x\},\\
\textup{THR}(x)&=\P\{Y(0)=1,Y(1)=0\mid X=x\},\\
\textup{TNR}(x)&=\P\{Y(1)=Y(0)\mid X=x\}.
\end{split}
\]
Then $\tau(x)=\textup{TBR}(x)-\textup{THR}(x)$. We use the CATE here only as a benchmark. By contrast, these three conditional response rates depend on the joint distribution of $(Y(0),Y(1))$ and belong to the first sublayer of the third layer.

For a concrete illustration, suppose that $X$, $A$, and $Y$ are binary, with $Y=1$ denoting survival. In both subgroups, suppose that $\P\{Y(0)=1\mid X=x\}=0.4$ and $\P\{Y(1)=1\mid X=x\}=0.6$. The two subgroups therefore have the same marginal potential-outcome distributions and the same CATE, $\tau(x)=0.2$. In the subgroup with $X=0$, suppose that 20\% benefit, none are harmed, 40\% survive under either treatment, and 40\% die under either treatment. In the subgroup with $X=1$, suppose that 50\% benefit, 30\% are harmed, 10\% survive under either treatment, and 10\% die under either treatment. Both joint response distributions are compatible with the same two marginals, but their benefit--harm profiles differ sharply. Thus, it is the joint response-type distribution, rather than the CATE itself, that supplies the third-layer information. This distribution can matter when the scientific or treatment decision concerns the subgroup balance of benefit and harm, although it does not reveal any particular individual's unobserved response type.
\end{example}

\begin{example}[Effect of Persuasion] \label{example5}
Let $A \in \{0, 1\}$ denote a binary indicator of an individual’s exposure to persuasive information, and let 
$Y$ be a binary outcome, where $Y=0$ indicates a negative response and $Y=1$ indicates a positive response.  The persuasion rate~\citep{Jun-Lee2023, Jun-Lee2024} is defined as  
$$\textup{PR} = \mathbb{P}\!\left( Y(1) = 1 \mid Y(0) = 0 \right),$$
which  
quantifies the proportion of individuals who would respond negatively in the absence of exposure but whose behavior becomes positive after exposure to persuasive information.
\end{example}

\begin{example}[Distribution of ITE] \label{example6}
For binary treatments, we can define the distribution of the ITE as $F(\delta) = \P(Y(1) - Y(0) \leq \delta)$, as studied by \citet{Kim2014, Yin-etal2018, Shin2025}. This distribution differs from the $\textup{DTE}(y)$ introduced in Example \ref{example1}. 
Several studies have considered variants of $F(\delta)$. For example,  
\citet{Kallus2023} defined the conditional value at risk (CVaR) of the ITE distribution, 
		\[     \textup{CVaR}_{\alpha} = \E[ Y(1) - Y(0)  \mid Y(1) - Y(0) \leq  q_{\alpha} ],     \]
where $q_{\alpha}$ is the $\alpha$-quantile of $Y(1) - Y(0)$. The $\textup{CVaR}_{\alpha}$ measures the average treatment effect in the lower $\alpha$-quantile of the treatment effect distribution. 
\citet{Kaji-Cao2025} studied estimands of the form $\E[ Y(1) - Y(0) \mid Y(0) \leq c ]$. If $Y$ denotes wealth, this represents the average treatment effect for the subgroup that would have low wealth if untreated.   
\end{example}

\subsubsection{Identifiability Challenges and Corresponding Strategies}
Identifying causal estimands at this layer requires characterizing the joint distribution of potential outcomes. Because both potential outcomes are never observed simultaneously for any individual, this task poses substantial identification challenges.  
Beyond confounding between treatment and outcome, one must additionally characterize the dependence structure between potential outcomes. Randomization resolves the former, but not the latter, and therefore does not generally identify third-layer estimands.
Below, we review various strategies for addressing this dependence in the case of a binary treatment. 



 $\bullet$ \emph{Independence Between Potential Outcomes: $Y(1) \indep Y(0)$}.    
Under this assumption, the joint distribution of potential outcomes $(Y(1), Y(0))$ is determined by their 
marginal distributions. \citet{Yin-etal2018} relax this assumption to latent independence, $Y(1) \indep Y(0) \mid U$ for a latent variable $U$, at the cost of additional parametric model restrictions.  

 $\bullet$ \emph{Monotonicity Between Potential Outcomes: $Y(1) \geq Y(0)$ a.s.}.
Monotonicity addresses the dependence between $Y(1)$ and $Y(0)$ for binary outcomes and has been widely used in the literature; see, e.g., \citet{tian2000probabilities, cai2005variance, Huang-etal2012}. 
The implication of this condition is straightforward: under monotonicity and ignorability, the joint distribution $\mathbb{P}(Y(1), Y(0))$ involves three unknown parameters, $\pi_{jk} = \mathbb{P}(Y(1)=j, Y(0)=k)$ for $j,k \in {0,1}$ with $\pi_{01}=0$, which are identified by the system of three equations: $\pi_{10} + \pi_{11} = \mathbb{P}(Y(1)=1), 
\pi_{01} + \pi_{11} = \mathbb{P}(Y(0)=1), 
\sum_{j=0}^1 \sum_{k=0}^1 \pi_{jk} = 1.
$
Nevertheless, an interesting implication of the monotonicity assumption is an asymmetry in its identifying power. While monotonicity is sufficient to identify the joint distribution of potential outcomes when the treatment takes multiple ordered levels and the outcome is binary \citep{wang2017causal}, it generally does not yield point identification when outcomes are ordinal or continuous, even with binary treatment \citep{Zhang-etal2025, Lu-etal2025, Zhang-Yang2025}. 


 $\bullet$ \emph{Specification of Association Parameter Between Potential Outcomes}.    
Under the ignorability and overlap assumptions, for binary outcomes, the joint distribution of $(Y(1), Y(0))$ is identifiable for a given Pearson correlation coefficient  $\rho = \text{Corr}(Y(1), Y(0))$~
\citep{Wu-etal-2024-Harm}, or a given odds ratio~\citep{ciocuanea2025sensitivity,tong2025semiparametric}
  \begin{eqnarray*}\text{OR} = \frac{\P(Y(1)=1 \mid  Y(0)=1) \P(Y(1) =0 \mid  Y(0) = 0)}{\P(Y(1) =0 \mid  Y(0) = 1) \P(Y(1) = 1 \mid  Y(0) = 0)}. \end{eqnarray*}
When specifying a single association parameter is difficult, one may instead posit a plausible range for it and derive corresponding lower and upper bounds for the joint distribution.  This naturally yields a sensitivity analysis framework, with the association parameter serving as the sensitivity parameter.
A key insight of \citet{Wu-etal-2024-Harm} is that assuming a positive association parameter is often reasonable in real-world applications and can substantially tighten the bounds on the joint distribution of potential outcomes.  


 $\bullet$ \emph{Copula Models for Continuous Outcomes.} 
Specifying an association parameter alone is insufficient to identify the joint distribution of potential outcomes for continuous outcomes, necessitating additional model restrictions. Copula models offer a classical framework for constructing joint distributions from marginal distributions by explicitly modeling their dependence structure \citep{Jaworski-etal-compula2010, Bartolucci01062011, Sun-etal2024, Lu-etal2025, Zhang-Yang2025}. For example, a Gaussian copula assumes that $(Y(1), Y(0))$ follows a joint Gaussian distribution with a given correlation coefficient; varying this coefficient naturally yields a sensitivity analysis framework.



 $\bullet$ \emph{Data Fusion.} For binary and categorical outcomes, \citet{Wu-Mao2025} establishes nonparametric identification of the joint distribution under binary treatment by combining multiple experimental studies, and \citet{Shahn-Madigan2025} subsequently extends this framework.

Although researchers can use the above strategies to address the identification of the joint distribution of potential outcomes, the required conditions may be too restrictive in practice. Consequently, partial identification naturally arises as a central framework for third-layer causal inference. Rather than aiming for full recovery of the joint distribution of potential outcomes,  
partial identification methods characterize the set of distributions or provide bounds on the joint distribution of potential outcomes, compatible with the observed data and the maintained assumptions, see e.g., \citet{Fan-Park-2009, Fan-Park-2010, Fan-etal2014, Kim2014, Firpo2019, Frandsen2021, Kaji-Cao2025}.



One may notice that the applicability of the identification strategies reviewed above depends critically on the types of treatment and outcome variables. We briefly discuss this point in Remark \ref{rmk2}.

\begin{remark} \label{rmk2}  
Treatment type affects both the dimension of the joint potential outcome distribution and the plausibility of restrictions used to identify it. With a binary treatment, the coupling problem involves two potential outcomes. Ordered or multivalued treatments require a coherent joint distribution over several potential outcomes and an ordering under which monotonicity is scientifically meaningful. For unordered categorical or continuous treatments, a single pairwise monotonicity restriction generally does not determine the required dependence structure, so stronger structural modeling or partial identification may be required.

For binary outcomes, assumptions such as monotonicity, specification of an odds ratio, or a correlation parameter are often sufficient to fully determine the joint distribution of $(Y(1),Y(0))$ given the marginal distributions.
For continuous outcomes, however, a scalar association measure (e.g., Pearson correlation) does not identify the full joint distribution without further structural assumptions. Copula models or rank preservation assumptions are typically required to couple the two marginals. This substantially increases the modeling burden and sensitivity to specification. 
For ordered or multivalued categorical outcomes, monotonicity can be extended under suitable order restrictions, but more flexible models, such as latent threshold or multinomial probit formulations, may be needed to capture the dependence structure adequately.

For \textit{data fusion approaches} that combine multiple experimental studies, identification is most naturally achieved when outcomes are categorical with finite support; extending these methods to continuous outcomes poses additional challenges and remains an ongoing research direction.

We therefore encourage practitioners to select identification strategies that are compatible with the measurement scale of their variables and the plausibility of the required assumptions, and to interpret results with due caution when assumptions are strong.
\end{remark}

\subsection{Third Layer: Individual-Level Counterfactual Queries} \label{Sec3-4}
Different from Sections~\ref{Sec3-2}–\ref{Sec3-3}, this layer first requires clarifying the well-definedness of the task of learning individual-level counterfactual outcomes.

\subsubsection{Well-Definedness and Identifiability Challenges}

 From a probabilistic perspective, learning individual-level counterfactual outcomes is generally infeasible, as they are random variables rather than well-defined causal estimands. To address this issue, we need to impose the assumption below. 

\begin{assumption}[Deterministic Counterfactual Outcomes,  
\citeauthor{Pearl-etal2016-primer}, \citeyear{Pearl-etal2016-primer}] \label{assump3-1}
For the population of interest, we treat each individual's potential outcome $Y(a)$ as a fixed quantity. 
\end{assumption}

{Assumption~\ref{assump3-1} adopts a deterministic view of counterfactual outcomes, which differs from the stochastic perspective underlying statistical modeling and inference, where $Y(a)$ is treated as a random variable \cite{Dawid2000}.} 
This deterministic view is adopted primarily as an operational device for reasoning about individualized counterfactual outcomes. 
Under Assumption~\ref{assump3-1}, learning individual-level counterfactual outcomes becomes feasible under suitable conditions. Beyond confounding between treatment and outcome, the key challenge lies in establishing a connection between the factual and counterfactual outcomes at the individual level. 
Before presenting the identifiability strategies, we provide an example illustrating the risk of individual decision-making based on the conditional ATE (CATE), an estimand in the second layer. 

\begin{example}[Risks of Decision-Making Based on CATE]
In policy learning and personalized medicine, it is common to recommend treatment for an individual with $X=x$ if their CATE, $\tau(x) = \E[Y(1) - Y(0) \mid X=x]$, is greater than zero. However, $\tau(x)$ is an average metric over the subpopulation with $X=x$, and using it for individualized decision-making can be misleading~\citep{Ding-etal2019, Lei-Candes2021, MuellerPearl2023-CausalInference}. 
For example, consider a subpopulation with covariates $X=x$ consists of 10 individuals, and $\tau(x) = 0.1$. Based on CATE, we would recommend treatment for all individuals in this subpopulation. However, it may be the case that five individuals have ITEs of 1, while the other five have ITEs of -0.8. Recommending treatment for all 10 individuals would therefore harm half of the subpopulation. 
To address this problem and promote safer decision-making, we can use the harm rate as a complementary criterion for treatment recommendations~\citep{li2023trustworthy, 2022nathan, Wu-etal-2024-Harm}, or construct prediction intervals for the ITE~\citep{Jin-etal2023b}.

\end{example}

\subsubsection{Identifiability and Estimation Strategies} 

\citet{Pearl-etal2016-primer} proposed the widely used three-step procedure--abduction, action, and prediction--for estimating counterfactual outcomes, briefly summarized below.

\smallskip 
 \emph{\bf Pearl's Three-Step Procedure.} Suppose we have three sets of variables $A$, $Y$, and ${\bf E} \subseteq {\bf V}$ in a structural causal model (SCM) $\mathcal{M}$. Consider the counterfactual query: for an individual with observed evidence ${\bf E}={\bf e}$, what would have happened had we set $A$ to a different value $a'$? 
Pearl's three-step procedure addresses this question as follows:
(a) {\bf Abduction}: infer the value of the exogenous variables ${\bf U}$ for this individual given the evidence ${\bf E}={\bf e}$;
(b) {\bf Action}: modify $\mathcal{M}$ by replacing the structural equations for $A$ with $A=a'$, yielding the intervened model $\mathcal{M}_{a'}$;
(c) {\bf Prediction}: use $\mathcal{M}_{a'}$ together with the inferred ${\bf U}$ to compute the counterfactual outcome of $Y$ for this individual.

For clarity, we present an example illustrating the application of Pearl's three-step procedure. Specifically, let $\mathbf{V} = (X, A, Y)$, where $A$ causes $Y$, $X$ affects both $A$ and $Y$, and the structural equation of $Y$ is given as 
	\begin{equation}  \label{eq1}
   Y = f_Y(A, X, U).  
	 \end{equation} 
Let $Y(a')$ denote the potential outcome that would result if $A$ were set to $a'$.  
The counterfactual question, ``given the observed evidence $(A=a, X=x, Y=y)$ for an individual, what would have happened had $A$ been set to $a'$?'', corresponds to estimating $y_{a'}$, the realized value of $Y(a')$ for that individual. 
Using Pearl's three-step procedure, estimation of $y_{a'}$ proceeds as follows:   
(a) \emph{Abduction}: infer the individual-specific value of the exogenous variables, denoted by $u$;  
(b) \emph{Action}: intervene to set $A=a'$; and  
(c) \emph{Prediction}: compute the counterfactual outcome as $f_Y(a', x, u)$.

\smallskip
{\bf Analysis of Pearl's Three-Step Procedure.} 
Pearl's three-step procedure provides an elegant and easily manipulable method for estimating individual-level counterfactual outcomes. However, its application requires two prerequisites:
(a) The availability of an SCM that describes the data-generating process~\citep{Brouwer2022, Xie-etal2023-attribution}; and
(b) the exogenous variables must be identifiable or estimable from the SCM. For instance, in  model \eqref{eq1}, one needs to determine the functional form of $f_Y$ and obtain the value of $U$  using the inverse of $f_Y$.  
These two prerequisites may limit the applicability of Pearl's three-step procedure.

We further summarize two alternative methods--quantile regression and conformal inference--for estimating and bounding individual-level counterfactual outcomes, respectively. 

\smallskip 
\emph{\bf Rank Preservation and Quantile Regression}. 
\citet{Xie-etal2023-attribution} established the identifiability of counterfactual outcomes under ignorability and a strict monotonicity assumption, under which the outcome is a strictly monotone function of an exogenous variable. Under this assumption, for any individual, the factual and counterfactual outcomes share the same quantile in their respective marginal distributions of potential outcomes. 
For estimation, they proposed a quantile regression method to estimate counterfactual outcomes, thereby avoiding the need to specify or estimate a structural causal model.
\citet{Wu-etal2025-Rank} further showed that the strict monotonicity assumption is a special case of rank preservation, under which an individual’s factual and counterfactual outcomes share the same rank in the corresponding outcome distributions for all individuals~\citep{Heckman1997, Chernozhukov2005}, and proposed an improved estimation method.
\citet{wang2025counterfactually} used a similar technique to estimate fairness metrics, with applications to reinforcement learning.

From a distributional perspective, rank preservation implicitly specifies a particular coupling between the factual and counterfactual outcome distributions that coincides with the optimal transport map, which for one-dimensional distributions aligns corresponding quantiles. This connection places rank preservation within the broader framework of optimal transport, as, for example, studied for causal effects on outcome distributions by \citet{Lin-Kong-Wang2023}.

\smallskip 
 {\bf Conformal Inference}. Instead of estimating counterfactual outcomes for each individual, conformal inference aims to construct prediction intervals for them~\citep{Lei-Candes2021, Jin-etal2023b, Bodik2025}. For example, in the case of a binary treatment, conformal inference seeks to find prediction intervals $C_{\textup{ITE}}(X)$ for the individual treatment effect (ITE) such that, for a given
 $\alpha \in (0, 1)$, 
 \[   \P(  Y(1) - Y(0) \in C_{\textup{ITE}}(X; \alpha) ) \geq 1 - \alpha, \]
or $\P(  Y(1) - Y(0) \in C_{\textup{ITE}}(X; \alpha) \mid X=x ) \geq 1 - \alpha$. 
{For an individual $i$, without loss of generality, if $Y_i(0)$ is observed and $Y_i(1)$ is missing, we only need to construct prediction interval for $Y_i(1)$, i.e., find $C_1(X_i; \alpha)$ such that 
$
\P(Y_i(1) \in C_1(X_i; \alpha)) \geq \alpha.
$ 
It is then natural to define $C_{\textup{ITE}}(X_i; \alpha) = C_1(X_i; \alpha) - Y_i(0).$ 
Under the ignorability and overlap assumptions, constructing $C_1(X_i; \alpha)$ reduces to a standard conformal inference problem with covariate shift between $\P(X)$ and $\P(X\mid A=1)$~\citep{Lei-Candes2021}. If both $Y_i(0)$ and $Y_i(1)$ are missing, we can first construct prediction intervals $C_0(X_i; \alpha/2)$ and $C_1(X_i; \alpha/2)$ for $Y_i(0)$ and $Y_i(1)$, respectively, and then define
$
C_{\textup{ITE}}(X_i; \alpha)
= \{ z_1 - z_2: z_1 \in C_1(X_i; \alpha/2),\; z_2 \in C_0(X_i; \alpha/2) \}
$~\citep{Jin-etal2023b}.} 

The prediction intervals $C_{\text{ITE}}(X_i; \alpha)$ are often wide. \citet{Bodik2025} obtained narrower intervals by introducing a mild specification (positive correlation) of the association parameters between potential outcomes, extending the work of \citet{Wu-etal-2024-Harm}.

\medskip 
{\bf Comparison.} We compare the merits of rank preservation (quantile regression) and conformal inference methods, as summarized in Table~\ref{tab:method_comparison}. 

 \begin{table}[h]
\caption{Comparison of rank preservation and conformal inference.}
\label{tab:method_comparison}
\centering
\resizebox{0.65\columnwidth}{!}{\begin{tabular}{lcc}
\toprule
Property & Rank Preservation & Conformal Inference \\
\midrule
Weaker Conditions      & \textcolor{blue!70}{\Large \frownie}   &  \textcolor{orange!80!red}{\Large\smiley}  \\
Point Identification      &  \textcolor{orange!80!red}{\Large\smiley}   & \textcolor{blue!70}{\Large \frownie}  \\
Generalizable         & \textcolor{blue!70}{\Large \frownie}   &  \textcolor{orange!80!red}{\Large\smiley}  \\
\bottomrule
\end{tabular}}
\end{table}

Estimating individual-level counterfactual outcomes yields point estimates of the ITE, but this typically relies on strong identifiability assumptions (e.g., rank preservation). Moreover, such approaches are inherently retrospective and difficult to generalize, since counterfactual estimation depends on observed factual outcomes and therefore applies only to populations with observed outcomes.  
In comparison, conformal inference provides a promising framework for conducting inference on the ITE with high probability while relying on substantially weaker assumptions. Moreover, it exhibits strong generalization capability, in that it can be applied to new populations for which only covariates are observed. Nevertheless, the resulting prediction intervals are often wide and provide only limited information about the ITE, restricting their applicability.


\subsection{In the Presence of Post-Treatment Variable}  \label{Sec3-5}

In this subsection, we adopt the proposed potential outcomes perspective to briefly analyze scenarios in which an additional post-treatment variable is present. In addition to the observed variables $(A, X, Y)$, suppose that we also observe a variable $S$ measured after the treatment $A$ and before the outcome $Y$. 
Consider a binary treatment, and let $S(0)$ and $S(1)$ denote the potential outcomes of $S$.

\begin{example}[Principal Causal Effects] \label{example7}
The principal causal effect is defined as  
    $$\E[ Y(1) - Y(0) \mid S(1) = s_1, S(0) = s_0 ].$$
This estimand belongs to the third layer, as it involves the joint distribution of $(S(0), S(1))$.
The principal causal effect has many practical applications, by choosing different variable $S$ and specific principal strata $(s_1, s_0)$. For example,
in \textit{noncompliance} settings, let $S$ denote the actual treatment received; setting $(s_1,s_0)=(1,0)$ targets the complier average causal effect~\citep{Imbens-Angrist1994}. 
In \textit{truncation by death} settings, let $S$ be survival status; setting $(s_1,s_0)=(1,1)$ yields the survivor average causal effect~\citep{Rubin2006, wang2017identification}. 
In \textit{surrogate endpoint evaluation} settings, let $S$ be a biomarker; stratification on its joint potential values provides causal evidence for surrogacy \citep{Jiang-etal2016, Wu-Mao2025}. 
All these estimands are special cases of the principal causal effect under the same principal stratification framework. 
For additional examples, see \citet{Frangakis-Rubin2002, Lu-etal2025, Zhang-Yang2025}.


\end{example}

\begin{example}[Short-term and Long-term Treatment Effects] \label{example8} Let $S$ and $Y$ denote the short-term and long-term outcomes, respectively. Then $\E[S(1) - S(0)]$
and 
$\E[Y(1) - Y(0)]$ represent the short-term and long-term treatment effects, respectively, which belong to estimands in the second layer. 
A key characteristic of this setting is that long-term outcomes are difficult to observe and often suffer from substantial missingness due to extended follow-up periods, drop-outs, and budget constraints~\citep{athey-etal2019, chen2021semiparametric, kallus2024role, hu2023longterm, imbens2022long}. Beyond addressing confounding between treatment and both short- and long-term outcomes, one must also account for the missingness of long-term outcomes~\citep{athey-etal2019}. 
Several studies investigate how to learn policies that maximize long-term outcomes~\citep{Yang2024, Huang-etal2024}, or how to balance short- and long-term outcomes~\citep{Wu-etal-ShortLong, yang2024learning}. 
\end{example}

\begin{example}[Causal Fairness Metrics] \label{example9}
Counterfactual fairness has gained increasing attention in recent years~\citep{Davide-Bradic2024, Li-etal2025}. Let \(A\) denote a protected attribute (e.g., gender), \(Y\) an outcome or task-relevant attribute (e.g., grade), and \(S \in \{0,1\}\) the decision (e.g., admission) made by a machine learning algorithm.
The goal is to evaluate the fairness of the algorithm. Several counterfactual fairness metrics have been proposed:

 $\bullet$ Counterfactual parity~\citep{Mitchell-etal2021}:  $\P(S(1) = 1) = \P( S(0) = 1)$. It belongs to the second layer. 

 $\bullet$ Counterfactual fairness~\citep{kusner2017counterfactual}: $S_i(1) = S_i(0)$ for all individual $i \in \{1,..., N\}$. It belongs to the third layer (individual-level counterfactual outcomes).  

$\bullet$  Principal fairness~\citep{Imai2023, Li-etal2025}: $\P(S=1 \mid Y(1), Y(0), A) = \P(S=1 \mid Y(1), Y(0))$ and  principal counterfactual parity: $\P( S(0) = 1 \mid Y(1) = Y(0) ) = \P( S(1) = 1 \mid Y(1) = Y(0) )$. These quantities belong to the third layer, 
 as they involve the joint distribution of potential outcomes.  
\end{example}

\begin{example}[Mediation Analysis]  \label{example10}
We denote the mediator variable by $M$ instead of $S$ by convention. We introduce the potential outcome $Y(a, m)$ corresponding to treatment $A=a$ and $M = m$. \citet{Pearl2001} further considered the nested potential outcome $Y(a, M(a))$, which represents the hypothetical outcome if the treatment were set to level $a$ and the mediator were set to its potential level $M(a)$ under treatment $a$. The potential outcome $Y(a)$ is defined as $Y(a, M(a))$ for $a = 0, 1$ (referred to as the composition assumption; \citeauthor{Ding2023}, \citeyear{Ding2023}).

$\bullet$ The total effect $
\text{TE} = \E\big[ Y(1) - Y(0) \big].$ 
This estimand belongs to the second layer because it requires only the marginal distributions of the potential outcomes and does not involve cross-world potential outcomes.

$\bullet$ 
The natural direct effect
$\text{NDE} = \E\big[ Y(1, M(0)) - Y(0, M(0)) \big],$
and the natural indirect effect
$\text{NIE} = \E\big[ Y(1, M(1)) - Y(1, M(0)) \big].$
These two estimands belong to the third layer, since the quantity \(Y(1, M(0))\) is cross-world (nested) potential outcomes that involves treatment and mediator values arising from different hypothetical worlds.

$\bullet$  The controlled direct effect 
$\text{CDE}(m) = \E[ Y(1, m) - Y(0, m)]$. 
It belongs to the second layer because it contrasts the marginal distributions of $Y(1,m)$ and $Y(0,m)$ under two joint interventions on the treatment and mediator, without requiring a cross-world joint distribution. This classification is distinct from identification from observed data, which can follow, for example, under sequential ignorability ($A\indep Y(a,m)\mid X$, $M\indep Y(a,m)\mid (A,X)$; or, equivalently, $(A,M)\indep Y(a,m)\mid X$).
This definition also presupposes that setting the mediator to $m$ represents a well-defined and scientifically meaningful intervention. That premise may be questionable when the mediator is not directly manipulable or when different ways of changing it have different consequences. Thus, the second-layer classification of the CDE should not be interpreted as making it a generally preferable alternative to a natural effect.

$\bullet$ Separable effects define a different intervention-based estimand rather than a generally preferable substitute for natural direct and indirect effects. They begin with a decomposition of the treatment into components, say $(A_Y,A_M)$, that are hypothesized to act through distinct pathways, and compare outcomes under component-wise interventions such as $\E\{Y(a_Y,a_M)\}-\E\{Y(a_Y',a_M)\}$; see, for example, \citet{Stensrud-etal-2021-Separable}. Unlike the CDE, a separable effect intervenes on treatment components rather than fixing the mediator itself. Under our distributional convention, a contrast of marginal outcome distributions under these single-world component interventions belongs to the second layer. However, this interpretation requires the treatment decomposition and the component-wise interventions to be substantively meaningful. Whether the effect is identifiable from data in which only the original treatment was assigned depends on further scientific and identification assumptions. Separable effects can avoid a cross-world definition only when these requirements are met, and they should not be conflated with controlled direct effects.
\end{example}

Example \ref{example10} also illustrates a conceptual distinction. The third layer is needed when the scientific question, and hence the causal estimand, inherently involves cross-world comparisons, as for natural direct and indirect effects. The total effect requires only marginal potential-outcome distributions and belongs to the second layer. The CDE and separable effects also have second-layer representations when the corresponding mediator or component-wise interventions are well defined. This classification does not establish a practical preference: the CDE presupposes a scientifically meaningful intervention on the mediator, while separable effects require a substantively meaningful treatment decomposition. These estimands answer different scientific questions and are not universal substitutes for natural effects.

\subsection{Challenges and Limitations of Third-Layer Estimands} \label{Sec3-6}

Although third-layer estimands can provide information about cross-world relationships that is not contained in marginal potential-outcome distributions, they also introduce important challenges and limitations.

First, unlike second-layer estimands, third-layer estimands generally cannot be identified from randomized experiments alone. Their identification typically requires additional assumptions on the dependence between potential outcomes, cross-world relationships, structural causal models, or individual-level counterfactual mechanisms. These assumptions are fundamentally untestable from the observed data and therefore require substantive scientific justification. 

Second, inference for third-layer estimands is often considerably more sensitive to modeling assumptions. Different assumptions, such as monotonicity, copula models, association parameters, or rank preservation, may all be compatible with the observed data while implying markedly different joint distributions of potential outcomes and consequently different values of the target estimands. Sensitivity analysis therefore plays a substantially more important role than in conventional second-layer causal inference.

Third, point identification is frequently unattainable without imposing strong structural assumptions. In many applications, partial identification provides a more appropriate inferential framework by characterizing the range of values compatible with both the observed data and the maintained assumptions. Consequently, interval-valued conclusions, rather than point estimates, should often be regarded as the primary inferential target for third-layer causal questions.

Finally, although third-layer estimands can provide additional information about subgroup response distributions, they do not necessarily replace second-layer estimands. The two layers address different scientific questions. Average treatment effects remain the appropriate target for population-level policy evaluation, whereas third-layer estimands may be warranted when the scientific objective intrinsically involves attribution, cross-world dependence, fairness assessment, or other cross-world causal questions. Controlled direct and separable effects should not be treated as generally preferable alternatives to cross-world natural effects. A controlled direct effect presupposes that fixing the mediator is a well-defined and scientifically meaningful intervention, which may be questionable in many applications; separable effects require a substantively meaningful decomposition of the treatment into components. These estimands avoid cross-world definitions only when the corresponding interventions are well defined and answer the scientific question, and they should not be viewed as universal substitutes for natural effects. Even then, such estimands describe distributions or probabilities over units and do not by themselves reveal an individual's unobserved response type. Choosing between second- and third-layer estimands should therefore depend on the scientific objective, the plausibility of the required assumptions, and the desired level of causal interpretation.

Overall, the additional information provided by third-layer estimands comes at the cost of stronger assumptions and increased inferential uncertainty. Recognizing this trade-off is essential for selecting appropriate estimands and interpreting their conclusions in practical applications.


\section{Conclusion and Discussion} %
\label{conclusion}

In this paper, we recast Pearl's causal hierarchy in potential outcomes language and made its familiar information ordering operational at the level of causal estimands. Rather than proposing a new hierarchy, we provide a diagnostic that classifies a target by the probabilistic object it requires: marginal potential outcome distributions, a joint distribution or nested cross-world quantity, or individual-level counterfactual outcomes. Applying this diagnostic to a broad collection of estimands, including ambiguous cases, yields a practical taxonomy of the questions each estimand can and cannot answer.
The coupling perspective provides the corresponding synthesis of identification strategies. Second-layer estimands depend only on marginal potential outcome distributions, whereas cross-world third-layer estimands additionally require a coupling between these marginals, namely their joint distribution. Monotonicity restricts the admissible couplings, copula models parameterize the dependence structure, rank preservation imposes a deterministic coupling that aligns quantiles, and partial identification characterizes the feasible couplings consistent with the observed data. This formulation makes explicit which part of a target is learned from the data and which part is supplied or restricted by assumptions. 

Estimands in the second and third layers answer different scientific questions. Second-layer estimands, such as the ATE and QTE, are appropriate when the target is a feature of marginal potential-outcome distributions and are identifiable under standard ignorability and overlap conditions. Third-layer estimands are warranted when the question intrinsically concerns attribution, a joint or nested cross-world quantity, or an individual-level counterfactual outcome. These targets pose fundamentally more challenging identification problems. We reviewed a range of strategies for addressing those problems, including monotonicity assumptions, copula models, Pearl’s three-step procedure, conformal inference, and partial identification.

Looking ahead, we highlight several promising directions for future research, building on the perspective developed in this paper and the identification strategies reviewed herein. 

\emph{First, generalizing identification strategies for cross-world queries.} Most current identification strategies (e.g., monotonicity assumptions and copula models) are either restrictive or parametric.  Developing less-restrictive and data-adaptive methods to partially or point-identify the joint distribution of potential outcomes, using tools such as optimal transport, generative modeling, or flexible bounding approaches, remains a critical research frontier. 

\emph{Second, moving from prediction to decision under third-layer objectives.} While conformal inference provides prediction intervals for ITEs, these intervals are often wide. Future research could combine conformal prediction with various partial identification assumptions to produce tighter prediction intervals, thereby yielding improved individual-level treatment rules. 

\emph{Third, algorithmic fairness and the third layer.} Our analysis of fairness metrics (Example 10) shows a clear progression from second-layer (group-level) counterfactual parity to third-layer (individual-level) counterfactual fairness. A future direction is to design machine learning algorithms that can directly optimize third-layer fairness criteria under weaker assumptions, perhaps leveraging recent advances in conformal prediction or partial identification.

In summary, the proposed potential outcomes perspective serves as a practical guide from scientific question, to estimand, to required probabilistic object and identifying assumptions. It helps researchers diagnose which layer their question occupies, assess whether their assumptions are sufficient, and navigate the growing toolkit of identification strategies. The appropriate layer is therefore question-dependent: a third-layer target is warranted when the scientific objective genuinely requires joint, nested, or individualized counterfactual information that marginal potential-outcome distributions alone cannot provide.

\bibliographystyle{plainnat}
\bibliography{references}

@techreport{RichardsonRobins2013,
  author      = {Thomas S. Richardson and James M. Robins},
  title       = {Single World Intervention Graphs (SWIGs): A Unification of the Counterfactual and Graphical Approaches to Causality},
  institution = {Center for Statistics and the Social Sciences, University of Washington},
  number      = {128},
  year        = {2013},
  note        = {Working Paper}
}

@article{zhou2024promises,
  title={Promises of parallel outcomes},
  author={Zhou, Ying and Tang, Dingke and Kong, Dehan and Wang, Linbo},
  journal={Biometrika},
  volume={111},
  number={2},
  pages={537--550},
  year={2024},
  publisher={Oxford University Press}
}

@article{tang2026synthetic,
  title   = {The synthetic instrument: From sparse association to sparse causation},
  author  = {Tang, Dingke and Kong, Dehan and Wang, Linbo},
  journal = {Journal of the Royal Statistical Society Series B: Statistical Methodology},
  year    = {2026},
  volume    = {qkaf083}
}

@article{Wu-etal-2024-Harm,
	Author = {Peng Wu and Peng Ding and Zhi Geng and Yue Liu},
	Journal = {Journal of the Royal Statistical Society Series B: Statistical Methodology, In Press},
	Title = {Quantifying Individual Risk for Binary Outcomes},
	Year = {2026}}

@article{Ding-etal2019,
	author = {Peng Ding and Avi Feller and Luke Miratrix},
	journal = {Journal of the American Statistical Association},
	pages = {304--317},
	title = {Decomposing Treatment Effect Variation},
	volume = {114},
         number = {525}, 
	year = {2019}}

@article{Wu-Mao2025,
  title={The Promises of Multiple Experiments: Identifying Joint Distribution of Potential Outcomes},
  author={Peng Wu and Xiaojie Mao},
	Journal = {Journal of the Royal Statistical Society Series B: Statistical Methodology, In Press},
  year={2026}
}

@article{Dawid2022,
	author = {A. Philip Dawid and Monica Musio},
	journal = {Annual Review of Statistics and Its Application},
	pages = {261-287},
	title = {Effects of Causes and Causes of Effects},
	volume = {9},
	year = {2022}}

@inproceedings{2022nathan,
author = {Kallus, Nathan},
title = {What's the harm? Sharp bounds on the fraction negatively affected by treatment},
year = {2022},
booktitle = {Proceedings of the 36th International Conference on Neural Information Processing Systems},
publisher = {Curran Associates Inc.},
articleno = {1164},
numpages = {15996--16009},
series = {NIPS'22}
}

@article{shen2013treatment,
  title={Treatment benefit and treatment harm rate to characterize heterogeneity in treatment effect},
  author={Shen, Changyu and Jeong, Jaesik and Li, Xiaochun and Chen, Peng-Sheng and Buxton, Alfred},
  journal={Biometrics},
  volume={69},
  number={3},
  pages={724--731},
  year={2013},
  publisher={Oxford University Press}
}

@inproceedings{Luo-etal-2025,
author = {Luo, Shanshan and Yu, Yixuan and Liu, Chunchen and Xie, Feng and Geng, Zhi},
title = {Causal attribution analysis for continuous outcomes},
year = {2025},
publisher = {JMLR.org},
booktitle = {Proceedings of the 42nd International Conference on Machine Learning},
numpages = {41468--41493},
series = {ICML'25}
}

@inproceedings{li2023trustworthy,
author = {Li, Haoxuan and Zheng, Chunyuan and Cao, Yixiao and Geng, Zhi and Liu, Yue and Wu, Peng},
title = {Trustworthy policy learning under the counterfactual no-harm criterion},
year = {2023},
publisher = {JMLR.org},
booktitle = {Proceedings of the 40th International Conference on Machine Learning},
numpages = {20575--20598},
series = {ICML'23}
}

@article{Kallus2023,
    author = {Nathan Kallus},
    title = {Treatment Effect Risk: Bounds and Inference},
    journal = {Management Science},
    volume = {69},
    number = {8},
    pages = {4363--4971},
    year = {2023}
}

@article{Ding2023,
	author = {Peng Ding},
	journal = {arXiv:2305.18793},
	title = {A First Course in Causal Inference},
	year = {2023}}

@article{Shin2025,
	author = {Myungkou Shin},
	journal = {arXiv:2403.18503},
	title = {Distributional Treatment Effect with Latent Rank Invariance},
	year = {2025}}

@article{Kaji-Cao2025,
	author = {Tetsuya Kaji and Jianfei Cao},
	journal = {arXiv:2306.15048},
	title = {Assessing Heterogeneity of Treatment Effects},
	year = {2025}}

@article{wang2025counterfactually,
  title={Counterfactually Fair Reinforcement Learning via Sequential Data Preprocessing},
  author={Wang, Jitao and Shi, Chengchun and Piette, John D and Loftus, Joshua R and Zeng, Donglin and Wu, Zhenke},
  journal={arXiv preprint arXiv:2501.06366},
  year={2025}
}

@inproceedings{Brouwer2022,
  title     = {Deep counterfactual estimation with categorical background variables},
  author    = {Edward De Brouwer},
  booktitle = {Proceedings of the 36th International Conference on Neural Information Processing Systems},
  year      = {2022}, 
pages ={35213--35225},
publisher = {Curran Associates Inc.},
series = {NIPS'22}
}

@article{Kim2014,
	author = {Ju Hyun Kim},
	journal = {arXiv preprint arXiv:1410.5885},
	title = {Identifying the Distribution of Treatment Effects under Support Restrictions},
	year = {2014}}

@article{Firpo2019,
title = {Partial identification of the treatment effect distribution and its functionals},
journal = {Journal of Econometrics},
volume = {213},
number = {1},
pages = {210-234},
year = {2019},
author = {Sergio Firpo and Geert Ridder}
}

@article{Fan-etal2014,
author = {Yanqin Fan AND Robert Sherman AND Matthew Shum},
title = {Identifying Treatment Effects Under Data Combination},
journal = {Econometrica},
volume = {82},
number = {2},
pages = {811-822},
year = {2014}
}

@article{Frandsen2021,
author = {Brigham R. Frandsen and Lars J. Lefgren},
title = {Partial identification of the distribution of treatment effects with an application to the Knowledge is Power Program (KIPP)},
journal = {Quantitative Economics},
volume = {12},
number = {1},
pages = {143-171},
year = {2021}
}

@inproceedings{Li-etal2025,
author = {Li, Haoxuan and Tang, Zeyu and Jiang, Zhichao and Fang, Zhuangyan and Liu, Yue and Geng, Zhi and Zhang, Kun},
title = {Fairness on principal stratum: a new perspective on counterfactual fairness},
year = {2025},
publisher = {JMLR.org},
booktitle = {Proceedings of the 42nd International Conference on Machine Learning},
numpages = {35909--35921},
series = {ICML'25}
}

@article{Mitchell-etal2021,
	author = {Shira Mitchell and Eric Potash and Solon Barocas and Alexander D'Amour and Kristian Lum},
	journal = {Annual Review of Statistics and Its Application},
	pages = {141--163},
	title = {Algorithmic Fairness: Choices, Assumptions, and Definitions},
	volume = {8},
	year = {2021}}

@article{Fan-Park-2009,
	author = {Yanqin Fan and Sang Soo Park},
	journal = {Advances in Econometrics},
	pages = {3--70},
	title = {Partial Identification of the Distribution of Treatment Effects and its Confidence Sets},
	volume = {25},
       number = {3}, 
	year = {2009}}

@article{Fan-Park-2010,
	author = {Yanqin Fan and Sang Soo Park},
	journal = {Econometric Theory},
	pages = {931-951},
	title = {Sharp Bounds on the Distribution of the Treatment Effect and Their Statistical Inference},
	volume = {26},
       number = {3}, 
	year = {2010}}

@inproceedings{Ibeling-2020-AAAI,
	author = {Duligur Ibeling and Thomas Icard},
	booktitle = {AAAI Conference on Artificial Intelligence},
	title = {Probabilistic Reasoning across the Causal Hierarchy},
	year = {2020}}

@article{Firpo2007,
author = {Sergio Firpo},
title = {Efficient Semiparametric Estimation of Quantile Treatment Effects},
journal = {Econometrica},
volume = {75},
pages = {259-276},
number ={1}, 
year = {2007}}

@article{Pearl2019,
author = {Judea Pearl},
title = {The seven tools of causal inference, with reflections on machine learning},
journal = {Communications of the ACM},
volume = {62},
number = {3}, 
pages = {54-60},
year = {2019}}

@article{MuellerPearl2023-CausalInference,
	author = {Scott Mueller and Judea Pearl},
	journal = {Journal of Causal Inference},
	pages = {20220050},
	title = {Personalized Decision Making -- A Conceptual Introduction},
	volume = {11},
	year = {2023}}

@article{Chernozhukov2005,
author = {Victor Chernozhukov and Christian Hansen},
title = {An IV Model of Quantile Treatment Effects},
journal = {Econometrica},
volume = {73},
number = {1}, 
pages = {245-261},
year = {2005}}

@article{Heckman1997,
author = {James J. Heckman and Jeffrey Smith and Nancy Clements},
title = {Making the Most Out of Programme Evaluations and Social Experiments: Accounting
for Heterogeneity in Programme Impacts},
journal = {The Review of Economic Studies},
volume = {64},
pages = {487-535},
number ={5}, 
year = {1997}}

@article{Mueller-Pearl2023,
	author = {Aaron L Sarvet and Mats J Stensrud},
	journal = {American Journal of Epidemiology},
	title = {Perspective on `Harm' in Personalized Medicine},
	volume = {194},
	number = {6},
        pages = {1743--1748},
	year = {2023}}

@article{Stensrud-etal-2021-Separable,
  author = {Mats J. Stensrud and Miguel A. Hern{\'a}n and Eric J. Tchetgen Tchetgen and James M. Robins and Vanessa Didelez and Jessica G. Young},
  title = {A generalized theory of separable effects in competing event settings},
  journal = {Lifetime Data Analysis},
  year = {2021},
  volume = {27},
  number = {4},
  pages = {588--631},
  doi = {10.1007/s10985-021-09530-8}}

@article{Zhang-etal2013,
	author = {Zhiwei Zhang and Chenguang Wang and Lei Nie and Guoxing Soon},
	journal = {Journal of the Royal Statistical Society Series C: Applied Statistics},
	pages = {687--704},
	title = {Assessing the heterogeneity of treatment effects via potential outcomes of individual patients},
	volume = {62},
	number ={5}, 
	year = {2013}}

@article{Dawid2000,
	Author = {Dawid, A. P.},
	Journal = {Journal of the American Statistical Association},
	Title = {Causal Inference without Counterfactuals},
volume = {95},
number = {450},
pages = {407--424},
	Year = {2000}}

@article{Wu-etal-2025-Compare,
	Author = {Peng Wu and Shanshan Luo and Zhi Geng},
	Journal = {Journal of the American Statistical Association},
	Title = {On the Comparative Analysis of Average Treatment Effects Estimation via Data Combination},
volume = {120},
number = {552},
pages = {2250--2261},
	Year = {2025}}

@article{Gadbury2004,
	author = {Gary L. Gadbury and Hari K. Iyer and Jeffrey M. Albert},
	journal = {Journal of Statistical Planning and Inference},
	pages = {163-174},
	title = {Individual treatment effects in randomized trials with binary outcomes},
        number  = {2}, 
	volume = {121},
	year = {2004}}

@inproceedings{Wu-etal-ShortLong,
author = {Wu, Peng and Shen, Ziyu and Xie, Feng and Wang, Zhongyao and Liu, Chunchen and Zeng, Yan},
title = {Policy learning for balancing short-term and long-term rewards},
year = {2024},
publisher = {JMLR.org},
booktitle = {Proceedings of the 41st International Conference on Machine Learning},
pages = {53817--53846},
series = {ICML'24}
}

@inproceedings{yang2024learning,
author = {Yang, Qinwei and Liu, Xueqing and Zeng, Yan and Guo, Ruocheng and Liu, Yang and Wu, Peng},
title = {Learning the optimal policy for balancing short-term and long-term rewards},
year = {2024},
publisher = {Curran Associates Inc.},
booktitle = {Proceedings of the 38th International Conference on Neural Information Processing Systems},
pages = {36514--36540},
series = {NIPS'24}
}

@book{Pearl-Mackenzie2018,
	author = {Judea Pearl and Dana Mackenzie},
	publisher = {Hachette Book Group},
	title = {The Book of Why: The New Science of Cause and Effect},
	year = {2018}}

@article{Julian-etal2025,
	author = {Julian Dörfler and Benito van der Zander and Markus Bläser and Maciej Liskiewicz},
	journal = {arXiv preprint arXiv:2405.07373},
	title = {From Probability to Counterfactuals: the Increasing Complexity of Satisfiability in {P}earl's Causal Hierarchy},
	year = {2025}}

@article{wang2017identification,
  title={Identification and estimation of causal effects with outcomes truncated by death},
  author={Wang, Linbo and Zhou, Xiao-Hua and Richardson, Thomas S},
  journal={Biometrika},
  volume={104},
  number={3},
  pages={597--612},
  year={2017},
  publisher={Oxford University Press}
}

@article{Oka-etal2025,
	author = {Tatsushi Oka and Shota Yasui and Yuta Hayakawa and Undral Byambadalai},
	journal = {arXiv preprint arXiv:2407.14074},
	title = {Regression Adjustment for Estimating Distributional Treatment Effects in Randomized Controlled Trials},
	year = {2025}}

@article{Kennedy-etal2017,
    author = {Kennedy, Edward H. and Ma, Zongming and McHugh, Matthew D. and Small, Dylan S.},
    title = {Non-parametric Methods for Doubly Robust Estimation of Continuous Treatment Effects},
    journal = {Journal of the Royal Statistical Society Series B: Statistical Methodology},
    volume = {79},
    number = {4},
    pages = {1229-1245},
    year = {2017}
}

@article{Galvao02102015,
author = {Antonio F. Galvao and Liang Wang},
title = {Uniformly Semiparametric Efficient Estimation of Treatment Effects With a Continuous Treatment},
journal = {Journal of the American Statistical Association},
volume = {110},
number = {512},
pages = {1528--1542},
year = {2015},
publisher = {Taylor \& Francis}}

@book{Pearl-etal2016-primer,
	author = {Judea Pearl and Madelyn Glymour and Nicholas P. Jewell},
	publisher = {John Wiley \& Sons},
	title = {Causal Inference in Statistics: A Primer},
	year = {2016}}

@inproceedings{Xie-etal2023-attribution,
	author = {Shaoan Xie and  Biwei Huang and Bin Gu and Tongliang Liu and Kun Zhang},
	booktitle = {ICML Workshop
on Counterfactuals in Minds and Machines},
	title = {Advancing Counterfactual Inference through Quantile Regression},
	year = {2023}}

@article{imbens2022long,
	Author = {Imbens, Guido and Kallus, Nathan and Mao, Xiaojie and Wang, Yuhao},
	Journal = {Journal of the Royal Statistical Society Series B: Statistical Methodology},
	Title = {Long-term Causal Inference Under Persistent Confounding via Data Combination},
	volume = {87},
    number={2},
	pages = {362-388},
	Year = {2025}}

@article{wang2025causalinferencetaleframeworks,
	Author = {Linbo Wang and Thomas Richardson and James Robins},
	Journal = {arXiv preprint arXiv:2511.21516},
	Title = {Causal Inference: A Tale of Three Frameworks},
	Year = {2025}}

@article{dong2025marginalcausaleffectestimation,
	Author = {Mei Dong and Lin Liu and Dingke Tang and Geoffrey Liu and Wei Xu and Linbo Wang},
	Journal = {arXiv preprint arXiv:2510.14368},
	Title = {Marginal Causal Effect Estimation with Continuous Instrumental Variables},
	Year = {2025}}

@article{Jin-etal2023b,
	author = {Ying Jin and Zhimei Ren and Emmanuel J. Cand{\`e}s},
	journal = {Proceedings of the National Academy of Sciences},
	pages = {e2214889120},
	title = {Sensitivity analysis of individual treatment effects: A robust conformal inference approach},
	volume = {120},
	number ={6}, 
	year = {2023}}

@article{hu2023longterm,
  title={Identification and estimation of treatment effects on long-term outcomes in clinical trials with external observational data},
  author={Wenjie Hu and Xiao-Hua Zhou and Peng Wu},
  journal={Statistica Sinica},
	pages = {959-980},
	volume = {35},
  year={2025}
}

@article{Huang-etal2024,
author = {Huang, Ta-Wei and Ascarza, Eva},
title = {Doing More with Less: Overcoming Ineffective Long-Term Targeting Using Short-Term Signals},
year = {2024},
issue_date = {July-August 2024},
publisher = {INFORMS},
address = {Linthicum, MD, USA},
volume = {43},
number = {4},
journal = {Marketing Science},
pages = {863–884}
}

@article{Frangakis-Rubin2002,
  title={Principal Stratification in Causal Inference},
  author={Constantine E. Frangakis and Donald B. Rubin},
  journal={Biometrics},
  volume={58},
  number={1},
  pages={21--29},
  year={2002},
  publisher={International Biometric Society}
}

@article{Jiang-etal2016,
	author = {Jiang, Zhichao and Ding, Peng and Geng, Zhi},
	journal = {Journal of the Royal Statistical Society Series B: Statistical Methodology},
	month = {11},
	number = {4},
	pages = {829-848},
	title = {{Principal Causal Effect Identification and Surrogate end point Evaluation by Multiple Trials}},
	volume = {78},
	year = {2016}}

@article{Rubin2006,
author = {Donald B. Rubin},
title = {Causal Inference Through Potential Outcomes and Principal Stratification: Application to Studies with “Censoring” Due to Death},
volume = {21},
journal = {Statistical Science},
number = {3},
 year = {2006}, 
pages = {299--309}}

@article{Imbens-Angrist1994,
 author = {Guido W. Imbens and Joshua D. Angrist},
 journal = {Econometrica},
 number = {2},
 pages = {467--475},
 title = {Identification and Estimation of Local Average Treatment Effects},
 urldate = {2025-01-22},
 volume = {62},
 year = {1994}
}

@article{kallus2024role,
  title={On the role of surrogates in the efficient estimation of treatment effects with limited outcome data},
  author={Kallus, Nathan and Mao, Xiaojie},
  journal={Journal of the Royal Statistical Society Series B: Statistical Methodology},
  volume = {87}, 
  pages={480--509},
  year={2025},
  number = {2},
  publisher={Oxford University Press UK}
}

@article{Yang2024,
	Author = {Jeremy Yang and Dean Eckles and Paramveer Dhillon and Sinan Aral},
	Journal = {Management Science},
	Title = {Targeting for Long-Term Outcomes},
volume = {70},
number = {6},
pages = {3841–3855},
	Year = {2024}}

@article{chen2021semiparametric,
	Author = {Chen, Jiafeng and Ritzwoller, David M},
	journal = {Journal of Econometrics},
	volume = {237},
	number = {2},
        pages = {105545},
	Title = {Semiparametric estimation of long-term treatment effects},
	Year = {2023}}

@article{athey-etal2019,
	Author = {Athey, Susan and Chetty, Raj and Imbens, Guido and Kang, Hyunseung},
	Title = {The Surrogate Index: Combining Short-Term Proxies to Estimate Long-Term Treatment Effects More Rapidly and Precisely},
    journal = {The Review of Economic Studies},
    volume = {rdaf087},
	Year = {2025}}

@article{Bodik2025,
	Author = {Juraj Bodik and Yaxuan Huang and Bin Yu},
	Journal = {arXiv:2507.12581},
	Title = {Cross-World Assumption and Refining Prediction Intervals for Individual Treatment Effects},
	Year = {2025}}

@inproceedings{Pearl2001,
author = {Pearl, Judea},
title = {Direct and indirect effects},
year = {2001},
publisher = {Morgan Kaufmann Publishers Inc.},
booktitle = {Proceedings of the 17th Conference on Uncertainty in Artificial Intelligence},
pages = {411–420},
series = {UAI'01}
}

@article{Shpitser-Pearl2008,
author = {Shpitser, Ilya and Pearl, Judea},
title = {Complete Identification Methods for the Causal Hierarchy},
year = {2008},
volume = {9},
journal = {Journal of Machine Learning Research},
pages = {1941–1979},
numpages = {39}
}

@inbook{Bareinboim-etal2022,
author = {Bareinboim, Elias and Correa, Juan D. and Ibeling, Duligur and Icard, Thomas},
title = {On Pearl’s Hierarchy and the Foundations of Causal Inference},
year = {2022},
publisher = {Association for Computing Machinery},
address = {New York, USA},
edition = {1},
booktitle = {Probabilistic and Causal Inference: The Works of Judea Pearl},
pages = {507–556},
numpages = {50}
}

@article{Yin-etal2018,
	author = {Yunjian Yin and Lan Liu and Zhi Geng},
	journal = {Statistica Sinica},
	pages = {115-135},
	title = {Assessing the Treatment Effect Heterogeneity with a Latent Variable},
	volume = {28},
	year = {2018}}

@article{tong2025semiparametric,
  title={Semiparametric principal stratification analysis beyond monotonicity},
  author={Tong, Jiaqi and Kahan, Brennan and Harhay, Michael O and Li, Fan},
  journal={arXiv preprint arXiv:2501.17514},
  year={2025}
}

@article{Bartolucci01062011,
author = {Francesco Bartolucci and Leonardo Grilli},
title = {Modeling Partial Compliance Through Copulas in a Principal Stratification Framework},
journal = {Journal of the American Statistical Association},
volume = {106},
number = {494},
pages = {469--479},
year = {2011}
}

@article{Davide-Bradic2024,
author = {Viviano Davide and Jelena Bradic},
title = {Fair Policy Targeting},
journal = {Journal of the American Statistical Association},
volume = {119},
number = {545},
pages = {730--743},
year = {2024}
}

@article{Shahn-Madigan2025,
	author = {Zach Shahn and David Madigan},
	journal = {arXiv preprint arXiv:2509.20506},
	title = {Identification and Estimation of Joint Potential Outcome Distributions from a Single Study},
	year = {2025}}

@inproceedings{Wu-etal2025-Rank,
    author = {Peng Wu and Haoxuan Li and Chunyuan Zheng and Yan Zeng and Jiawei Chen and Yang Liu and Ruocheng Guo and Kun Zhang},
  title     = {Learning Counterfactual Outcomes Under Rank Preservation},
booktitle = {Proceedings of the 39th International Conference on Neural Information Processing Systems}, 
  year      = {2025},
publisher = {Curran Associates Inc.}, 
  series = {NIPS'25}
}

@book{Jaworski-etal-compula2010,
	author = {P. Jaworski and F. Durante and W. K. Hardle and T. Rychlik},
	publisher = {Springer},
	title = { Copula Theory and Its Applications},
	volume = {198},
	year = {2010}}

@article{Sun-etal2024,
	Author = {Shuo Sun and Johanna G. Nešlehová and Erica E. M. Moodie},
	Journal = {Statistics in Medicine},
	Pages = {34--38},
	Publisher = {Wiley Online Library},
	Title = {Principal stratification for quantile causal effects under partial compliance},
	Volume = {43},
	number = {1}, 
	Year = {2024}}

@article{Lu-etal2025,
    author = {Lu, Sizhu and Jiang, Zhichao and Ding, Peng},
    title = {Principal stratification with continuous post-treatment variables: nonparametric identification and semiparametric estimation},
    journal = {Journal of the Royal Statistical Society Series B: Statistical Methodology},
    volume = {88},
     number = {1}, 
    pages = {239--260},
    year = {2025}}

@article{ciocuanea2025sensitivity,
  title={Sensitivity analysis for the probability of benefit in randomized controlled trials with a binary treatment and a binary outcome},
  author={Cioc{\u{a}}nea-Teodorescu, Iuliana and Gabriel, Erin E and Sj{\"o}lander, Arvid},
  journal={Biostatistics},
  volume={26},
  number={1},
  pages={kxaf011},
  year={2025},
  publisher={Oxford University Press}
}

@article{Zhang-Yang2025,
	author = {Yichi Zhang and Shu Yang},
	journal = {Journal of the Royal Statistical Society Series B: Statistical Methodology},
	pages = {1655--1677},
	title = {Semiparametric localized principal stratification analysis with continuous strata},
	number = {5},
	volume = {87},
	year = {2025}}

@article{cai2005variance,
  title={Variance estimators for three “probabilities of causation”},
  author={Cai, Zhihong and Kuroki, Manabu},
  journal={Risk Analysis: An International Journal},
  volume={25},
  number={6},
  pages={1611--1620},
  year={2005},
  publisher={Wiley Online Library}
}

@article{tian2000probabilities,
  title={Probabilities of causation: Bounds and identification},
  author={Tian, Jin and Pearl, Judea},
  journal={Annals of Mathematics and Artificial Intelligence},
  volume={28},
  number={1},
  pages={287--313},
  year={2000},
  publisher={Springer}
}

@article{Huang-etal2012,
	author = {Ying Huang and Peter B. Gilbert and Holly Janes},
	journal = {Biometrics},
	pages = {687--696},
	title = {Assessing Treatment-Selection Markers using a Potential Outcomes Framework},
	volume = {68},
	year = {2012}}

@article{Jun-Lee2023,
  title={Identifying the effect of persuasion},
  author={Sung Jae Jun and Sokbae Lee},
  journal={Journal of Political Economy},
  volume={131},
  number={8},
  pages={2032--2058},
  year={2023}
}

@article{Jun-Lee2024,
  title={Learning the Effect of Persuasion via Difference-In-Differences},
  author={Sung Jae Jun and Sokbae Lee},
  journal={arXiv preprint arXiv:2410.14871},
  year={2024}
}

@article{Tian-Wu2025,
	author = {Zhaoqing Tian and Peng Wu},
	journal = {Statistics in Medicine},
	number = {18-19},
	pages = {e70242},
	title = {Semiparametric Efficient Inference for the Probability of Necessary and Sufficient Causation},
	volume = {44},
	year = {2025}}

@article{Lei-Candes2021,
	author = {Lihua Lei and Emmanuel J. Cand{\`e}s},
	journal = {Journal of the Royal Statistical Society Series B: Statistical Methodology},
	pages = {911-938},
	title = {Conformal Inference of Counterfactuals and Individual Treatment Effects},
	volume = {83},
	number = {5}, 
	year = {2021}}

@book{Imbens-Rubin2015,
	author = {G. W. Imbens and D. B. Rubin},
	publisher = {Cambridge University Press},
	title = {Causal Inference For Statistics Social and Biomedical Science},
	year = {2015}}

@book{Hernan-Robins2020,
	author = {M.A. Hern{\'a}n and J. M. Robins},
	publisher = {Boca Raton: Chapman and Hall/CRC},
	title = {Causal Inference: What If},
	year = {2020}}

@article{Colnet-etal2024,
	author = {B{\'e}n{\'e}dicte Colnet and Imke Mayer and Guanhua Chen and Awa Dieng and Ruohong Li and Ga{\"e}l Varoquaux and Jean-Philippe Vert and Julie Josse and Shu Yang},
	journal = {Statistical Science},
	title = {Causal inference methods for combining randomized trials and observational studies: a review},
	volume = {39},
         number = {1}, 
    	pages = {165-191},
	year = {2024}}

@article{Imbens2004,
    author = {Imbens, Guido W.},
    title = {Nonparametric Estimation of Average Treatment Effects Under Exogeneity: A Review},
    journal = {The Review of Economics and Statistics},
    volume = {86},
    number = {1},
    pages = {4-29},
    year = {2004}}

@book{Rosenbaum2020,
	Author = {Paul R. Rosenbaum},
	Publisher = {Springer Series in Statistics},
	Title = {Design of Observational Studies},
	Year = {2020}}

@inproceedings{Ding-etal2022,
author = {Ding, Sihao and Wu, Peng and Feng, Fuli and Wang, Yitong and He, Xiangnan and Liao, Yong and Zhang, Yongdong},
title = {Addressing Unmeasured Confounder for Recommendation with Sensitivity Analysis},
year = {2022},
publisher = {Association for Computing Machinery},
booktitle = {Proceedings of the 28th ACM SIGKDD Conference on Knowledge Discovery and Data Mining},
pages = {305–315},
series = {KDD'22}
}

@inproceedings{Kallus-Zhou2018,
author = {Kallus, Nathan and Zhou, Angela},
title = {Confounding-robust policy improvement},
year = {2018},
publisher = {Curran Associates Inc.},
booktitle = {Proceedings of the 32nd International Conference on Neural Information Processing Systems},
pages = {9289–9299},
numpages = {11},
series = {NIPS'18}
}

@article{Lipsitch-etal2010,
    author = {Marc Lipsitch and Eric Tchetgen Tchetgen and Ted Cohen},
    title = {Negative controls: a tool for detecting confounding and bias in observational studies},
    journal = {Epidemiology},
    volume = {21},
 number = {1},
    pages = {383-388},
    year = {2010}}

@article{wang2018bounded,
  title={Bounded, efficient and multiply robust estimation of average treatment effects using instrumental variables},
  author={Wang, Linbo and Tchetgen Tchetgen, Eric},
  journal={Journal of the Royal Statistical Society Series B: Statistical Methodology},
  volume={80},
  number={3},
  pages={531--550},
  year={2018},
  publisher={Oxford University Press}
}

@article{newey1993efficiency,
  title   = {Efficiency of weighted average derivative estimators and index models},
  author  = {Newey, Whitney K. and Stoker, Thomas M.},
  journal = {Econometrica},
  volume  = {61},
  number  = {5},
  pages   = {1199--1223},
  year    = {1993}
}

@article{Lin-Kong-Wang2023,
  title   = {Causal inference on distribution functions},
  author  = {Lin, Zheng and Kong, Dehan and Wang, Linbo},
  journal = {Journal of the Royal Statistical Society Series B: Statistical Methodology},
  year    = {2023},
  volume  = {85},
  number  = {2},
  pages   = {378–-398}
}

@article{wang2017causal,
  title={Causal analysis of ordinal treatments and binary outcomes under truncation by death},
  author={Wang, Linbo and Richardson, Thomas S and Zhou, Xiao-Hua},
  journal={Journal of the Royal Statistical Society Series B: Statistical Methodology},
  volume={79},
  number={3},
  pages={719--735},
  year={2017},
  publisher={Oxford University Press}
}

@article{pearl1999,
  author={Judea Pearl},
  title={Probabilities of causation: three counterfactual interpretations and their identification},
journal={Synthese},
  volume={121},
  pages={93--149},
  year={1999}
}

@article{lu2023evaluating,
  title={Evaluating causes of effects by posterior effects of causes},
  author={Lu, Zitong and Geng, Zhi and Li, Wei and Zhu, Shengyu and Jia, Jinzhu},
  journal={Biometrika},
  volume={110},
  number={2},
  pages={449--465},
  year={2023},
  publisher={Oxford University Press}
}

@inproceedings{li2024probabilities,
author = {Li, Ang and Pearl, Judea},
title = {Probabilities of causation with nonbinary treatment and effect},
year = {2024},
publisher = {AAAI Press},
booktitle = {Proceedings of the 38h AAAI Conference on Artificial Intelligence},
numpages = {20465--20472},
series = {AAAI'24}
}

@inproceedings{zhao2023conditional,
  title={Conditional counterfactual causal effect for individual attribution},
  author={Zhao, Ruiqi and Zhang, Lei and Zhu, Shengyu and Lu, Zitong and Dong, Zhenhua and Zhang, Chaoliang and Xu, Jun and Geng, Zhi and He, Yangbo},
booktitle = {Proceedings of the 39th Conference on Uncertainty in Artificial Intelligence},
  pages={2519--2528},
  year={2023},
series = {UAI'23}
}

@article{Zhang-etal2025,
    author = {Zhang, Chao and Geng, Zhi and Li, Wei and Ding, Peng},
    title = {Identifying and bounding the probability of necessity for causes of effects with ordinal outcomes},
    journal = {Biometrika},
    volume = {112},
    number = {3},
    pages = {asaf049},
    year = {2025}}

@article{li2024retrospective,
  title={Retrospective causal inference with multiple effect variables},
  author={Li, Wei and Lu, Zitong and Jia, Jinzhu and Xie, Min and Geng, Zhi},
  journal={Biometrika},
  volume={111},
  number={2},
  pages={573--589},
  year={2024},
  publisher={Oxford University Press}
}

@article{Hu-etal2023-NegativeControl,
    author = {Jie Kate Hu and Eric J. Tchetgen Tchetgen and Francesca Dominici },
    title = {Using negative controls to adjust for unmeasured confounding bias in time series studies},
    journal = {Nature Reviews Methods Primers},
    volume = {3},
    pages = {66},
    year = {2023}}

@article{rosenbaum1983central,
  title={The central role of the propensity score in observational studies for causal effects},
  author={Rosenbaum, Paul R and Rubin, Donald B},
  journal={Biometrika},
  volume={70},
  number={1},
  pages={41--55},
  year={1983},
  publisher={Oxford University Press}
}

@inproceedings{kusner2017counterfactual,
author = {Kusner, Matt and Loftus, Joshua and Russell, Chris and Silva, Ricardo},
title = {Counterfactual fairness},
year = {2017},
publisher = {Curran Associates Inc.},
booktitle = {Proceedings of the 31st International Conference on Neural Information Processing Systems},
pages = {4069–4079},
numpages = {11},
series = {NIPS'17}
}

@article{Imai2023,
	author = {Kosuke Imai and Zhichao Jiang},
	journal = {Statistical Science},
	title = {Principal Fairness for Human and
Algorithmic Decision-Making},
	volume = {38},
    number = {2},
	pages = {317--328},
	year = {2023}}

@book{pearl2009causality,
	author = {Pearl, Judea},
	publisher = {Cambridge University Press},
	title = {Causality},
	year = {2009}}

\end{document}